\documentclass[
twocolumn,
]{ceurart}

\usepackage{listings}

\usepackage{tabularx}
\usepackage{multicol}
\usepackage{subcaption}

\begin{document}

\copyrightyear{2022}
\copyrightclause{Copyright for this paper by its authors.
  Use permitted under Creative Commons License Attribution 4.0
  International (CC BY 4.0).}

\conference{QuASoQ`22: 10th International Workshop on Quantitative Approaches to Software Quality}

\title{Exploring the Impact of Code Style in Identifying Good Programmers}





\author[1]{Rafed Muhammad Yasir}[%
email=bsse0733@iit.du.ac.bd,
url=https://rafed.github.io/,
]
\address[1]{Institute of Information Technology (IIT),
  University of Dhaka, Dhaka, Bangladesh}

\author[1]{Dr. Ahmedul Kabir}[%
email=kabir@iit.du.ac.bd
]


\begin{abstract}
Code style is an aesthetic choice exhibited in source code that reflects programmers' individual coding habits. This study is the first to investigate whether code style can be used as an indicator to identify good programmers. Data from Google Code Jam was chosen for conducting the study. A cluster analysis was performed to find whether a particular coding style could be associated with good programmers. Furthermore, supervised machine learning models were trained using stylistic features and evaluated using recall, macro-F1, AUC-ROC and balanced accuracy to predict good programmers. The results demonstrate that good programmers may be identified using supervised machine learning models, despite that no particular style groups could be attributed as a good style.

\end{abstract}

\begin{keywords}
  code style \sep
  identify good programmer \sep
  stylistic features
\end{keywords}

\maketitle

\section{Introduction}
\begin{figure}[h]
\captionsetup{justification=centering}
\begin{multicols}{2}
    \includegraphics[width=\linewidth]{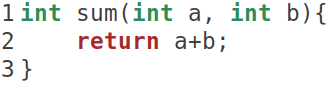}\par
    \vfill
    \subcaption{Style 1}
    
    \includegraphics[width=\linewidth]{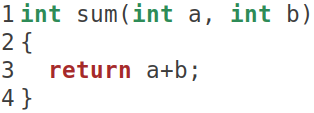}\par
    \subcaption{Style 2}
\end{multicols}
\caption{Two functionally same code snippets written in different styles}
\label{img:styleexample}
\end{figure}

Code style represents the physical layout of code (e.g., indentation, bracket placement), which reflects an individual's personal programming habits that do not affect its functionality \cite{ogura2018bring}. Figure \ref{img:styleexample} shows two code snippets that are functionally similar but written in two different styles. Code style has an impact on various aspects of software engineering, including software maintenance \cite{mi2016measuring} and speed of software development \cite{zou2019does}. However, no prior studies have been conducted to see whether good programmers can be detected by looking at their coding style. This paper investigates the potential for using code style to identify good programmers.

Establishing a link between code style and good programmers can have several implications. Many software repositories contain style guidelines that are used to enforce a specific code style in order to maintain software quality \cite{erkkinen2005model}. However, these style guides are often opinionated and arbitrary \cite{pullrequest, googlestyle}. If a specific code style exhibited by programmers can be identified as a good style, it can be used to create non-arbitrary style guidelines for better software maintenance.

In the software industry, the developers hired by a company directly affect  the quality of the codebase that they maintain. During recruitment, the candidates who apply for jobs often have to solve a set of programming problems. However, existing hiring practices do not account for the possibility that a skilled programmer could have a bad day and fail to answer a question correctly. Thus, in some circumstances a judgment may be unfair. If positive stylistic features can be identified in a programmer's code, they can be used as an additional criterion to enhance recruitment processes. This study is an initial attempt to determine whether such relationships between competent programmers and their code style can be established.


To conduct the study, the solutions collected from Google Code Jam (GCJ) \cite{gcjwebsite} were used as the dataset. 30 stylistic metrics were extracted from the codes and used as features for analysis. Two methods of analysis were used. At first, clustering algorithms were applied to the data to discover style groups and check whether good programmers belonged to a particular style group. Secondly, supervised machine learning models were trained using stylistic features to predict good programmers. The models were evaluated using recall, macro-F1, area under curve of ROC (AUC-ROC), and balanced accuracy.

Results show that, although style groupings were found, there were no specific groups with which good programmers could be associated. However, supervised machine learning models showed that good programmers can be predicted to some extent. Based on the evaluated metrics, a Balanced Random Forest achieved the best result with an average of 0.65 recall, 0.51 macro-F1, and 0.69 AUC-ROC.

\section{Related Work}

To the best of our knowledge, this is the first time code style has been used to identify good programmers. Early research conducted by Oman and Cook proposed a taxonomy for code styles to help people grasp a coherent view on the basis and application of code styles \cite{oman1991programming}. The four major categories of their taxonomy are general practices, typographic style, control structure style and information structure style. They also concluded in further research that code style is more than cosmetic and that it can affect areas such as code comprehension \cite{oman1990typographic}.

Caliskan et al. proposed a Code Stylometry Feature Set (CSFS) with which they performed source code authorship attribution \cite{caliskan2015anonymizing}. Their feature set is language agnostic and can be used for other programming languages as well. With their method, they achieved 94\% accuracy in classifying 1600 authors and 98\% accuracy in classifying 250 authors. They concluded that this method can help in the identification of authors of malicious programs, ghostwriting detection, software forensics and copyright investigation.

Mirza and Cosma explored the suitability of using code style in detecting plagiarism in the BlackBox dataset \cite{mirza2017style}. BlackBox is a project that collects data from users of the BlueJ which is a Java IDE \cite{mirza2017style}. Their study showed that code style is suitable for detecting plagiarism.

For evaluating software projects, Zou et al. explored how code style inconsistency can affect pull request integration in projects on Github \cite{zou2019does}. By analyzing 117 public repositories, they concluded that code styles with specific criteria can influence both the acceptance of pull requests and the time required to merge a pull request.

Mi and Yu conducted a study on stylistic inconsistency in software projects \cite{mi2016measuring}. They proposed a collection of stylistic metrics for C++ projects and used these metrics to analyze small-scale Github projects. By using hierarchical agglomerative clustering they showed that stylistic differences exist between source files in a project. They concluded that, using the degree of stylistic inconsistency as a basis, code comprehensibility and software maintainability could be improved in the future.

Several tools have been developed that can check stylistic inconsistencies and help programmers improve code style. Ala-Mutka et al. developed \emph{style++} that helps students learn good C++ programming conventions \cite{ala2004supporting}. Mäkelä et al. developed \emph{Japroach} that checks whether Java programs have a particular style and if style related issues exist in them \cite{makela2004japroch}. Ogura et al. developed \emph{stylecoordinator} to decrease inconsistency and improve code readability \cite{ogura2018bring}.

\section{Methodology}

\subsection{Dataset Description}
The dataset for the study was made up of the solutions gathered from the Google Code Jam (GCJ) website \cite{gcjwebsite}. GCJ is an annual programming contest held by Google. GCJ is selected because its data is publicly available and it can somewhat resemble programming exams in recruitment processes. Professional programmers, students and amateurs from all around the world participate in GCJ. Therefore, not only does the dataset consist of source code from varying sources, but they also solve the same problem which makes comparative study possible. The contest consists of seven rounds, each progressively harder than the previous. The rounds are: Qualification round, Round 1A, Round 1B, Round 1C, Round 2, Round 3 and World Finals. We consider the programmers who reached at least Round 3 as "good" programmers because participating in this round requires passing the previous rounds with a large number of accepted solutions.


Although GCJ accepts solutions in many programming languages, C++ was selected as the preferred language for evaluation as it is more prevalent among participants and has the highest number of submissions. Each problem of the contest has two validation sets: a small input set and a large input set. A solution for the large validation set is a valid solution for the small input set, but not vice versa. For our analysis, the solutions from the small input were taken as it had more submissions and it would also be redundant if both solutions were taken. A small number of solutions were rejected as the language encoding consisted of non-standard characters.

The solutions to the contests held in 2015, 2016, and 2017 are chosen for experimentation. However, we only include solutions from Qualification Round, Round 2 and Round 3 for our dataset. Round 1A, Round 1B, and Round 1C are excluded because participating in these rounds are optional and thus lacks submissions from all programmers. Solutions from the World Finals are excluded because the number of finalists is too small to take into consideration for analysis. Table \ref{tab:numofparticipants} shows the number of participants in each round of the contests.

\begin{table}[]
\centering
\caption{Number of Participants in a Round}
\label{tab:numofparticipants}
\begin{tabular}{|l|r|r|r|}
\hline
                             & \multicolumn{1}{c|}{\textbf{2015}} & \multicolumn{1}{c|}{\textbf{2016}} & \multicolumn{1}{c|}{\textbf{2017}} \\ \hline
\textbf{Qualification round} & 10744                              & 11401                              & 11342                              \\ \hline
\textbf{Round 2}             & 1650                               & 1641                               & 1824                               \\ \hline
\textbf{Round 3}             & 266                                & 296                                & 286                                \\ \hline
\textbf{World Finals}        & 22                                 & 20                                 & 21                                 \\ \hline
\end{tabular}
\end{table}

From the collected data, layout and lexical stylistic features were extracted based on \cite{caliskan2015anonymizing}. Abstract syntax tree based features were omitted as these features are not within the control of a programmer. Furthermore, term frequency based features were also excluded as they largely depend on the corpus being used. Following these criteria, 30 stylistic features were extracted. The features are listed in Table \ref{tab:features}.

\begin{table*}[]
\caption{Feature Description of Dataset}
\label{tab:features}
\begin{tabular}{|p{3.5cm}|p{10.3cm}|}
\hline
\textbf{Feature}        & \textbf{Definition}                                                                                                                                                                \\ \hline
numTabs/length          & Number of tabs divided by file length in characters                                                                                                                                     \\ \hline
numSpaces/length        & Number of space characters divided by file length in characters                                                                                                                         \\ \hline
numEmptyLines/length    & Number of empty lines divided by file length in characters                                                                                                                              \\ \hline
whiteSpaceRatio         & Ratio between the number of whitespace characters (spaces, tabs, and newlines) and non-whitespace characters                                                                       \\ \hline
newLineBeforeOpenBrace  & Ratio between the number of code blocks preceded by a newline character and not preceded by a newline character                                                                  \\ \hline
tabsLeadLines           & Ratio between the number of lines preceded by a tab and not preceded by a tab                                                                                                    \\ \hline
avgLineLength           & Average length of each line                                                                                                                                                        \\ \hline
stdDevLineLength        & Standard deviation of the lengths of each line                                                                                                                                     \\ \hline
numkeyword/length       & Number of occurrences of keyword divided by file length in characters, where keyword is one of if, else, else-if, for, while, do, break, continue, switch, case (10 different features) \\ \hline
numTernary/length      & Number of ternary operators divided by file length in characters                                                                                                                        \\ \hline
numTokens/length        & Number of word tokens divided by file length in characters                                                                                                                              \\ \hline
numUniqueTokens/length  & Number of unique keywords used divided by file length in characters                                                                                                                     \\ \hline
numComments/length      & Number of comments divided by file length in characters                                                                                                                                 \\ \hline
numLineComments/length  & Number of line comments divided by file length in characters                                                                                                                            \\ \hline
numBlockComments/length & Number of comments divided by file length in characters                                                                                                                                 \\ \hline
numLiterals/length      & Number of string, character, and numeric literals divided by file length in characters                                                                                                  \\ \hline
numMacros/length        & Number of preprocessor directives divided by file length in characters                                                                                                                  \\ \hline
nestingDepth            & Highest depth of control statements and loops                                                                                                                                      \\ \hline
numFunctions/length     & Number of functions divided by file length in characters                                                                                                                                \\ \hline
avgParams               & Average number of parameters of functions                                                                                                                                          \\ \hline
stdDevNumParams         & Standard deviation of the number of parameters of functions                                                                                                                        \\ \hline
\end{tabular}
\end{table*}

\subsection{Approach}
This section discusses the setups for exploring the effects of code style on classifying good programmers. Two methods were used for this purpose: (1) clustering techniques and (2) supervised machine learning algorithms.

\subsubsection{Analyzing Using Clustering Techniques}

Clustering is a method of partitioning objects into homogeneous groups on the basis of similarity among those objects \cite{johnson1967hierarchical}. t-SNE is one such algorithm that can discover the potential number of clusters in a dataset with high dimensions \cite{linderman2019clustering}. For each problem in the contests, t-SNE graphs were plotted with the intent of finding groups that conform to a particular style. Each data point in the plots represents a solution submitted by a programmer. The data points are labeled as:
\begin{itemize}
    \item Red: reached World Finals
    \item Green: reached Round 3
    \item Light blue: other programmers
\end{itemize}
The plots provide an estimate for the number of clusters and the distribution of good programmers in the clusters. 

To further validate the clustering provided by t-SNE, Hierarchical Agglomerative Clustering (HAC) with Ward linkage was performed and dendrograms were plotted. HAC is a clustering algorithm that treats every data point as a cluster and they are gradually merged to form a single cluster \cite{sasirekha2013agglomerative}. The number of clusters indicated by the dendrograms was matched with the number of clusters indicated by t-SNE before further analysis was performed.

To analyze the properties of the t-SNE clusters, solutions to each problem in the dataset were clustered using K-Means With K=number of clusters estimated by t-SNE. The solutions in the data were then labeled based on the cluster they belonged to. This labeled data was fitted to a Random Forest Classifier to obtain the feature importance of the tree. Based on the tree's feature importance, it was determined what style groups exist and whether good programmers belong to a specific style group.

\subsubsection{Analyzing using Supervised Machine Learning Algorithms}

Supervised learning is a method of training a model that can make predictions based on labeled data \cite{russell2016artificial}. For predicting good programmers, the following models were trained: Logistic Regression (LR), Support Vector Classifier (SVC), K-Nearest Neighbors (KNN), Decision Tree (DT) and Random Forest (RF). A Dummy classifier was also trained to act as a performance baseline for comparison \cite{dummyClassifier}. The models were trained for each problem in the dataset. Table \ref{tab:numofparticipants} shows that the number of participants in Round 3 is far less than the participants in the previous rounds. That is, the proportion of "good" programmers in the dataset is much lower in comparison to the other programmers. This makes the classification an imbalanced classification problem \cite{japkowicz2002class}. To balance the training data, the up-sampling technique SMOTE \cite{chawla2002smote} was used prior to training the above mentioned models. Furthermore, Balanced Random Forest (BRF) and RUS Adaboost classifier (RUSAda) were also trained which performs under-sampling to balance training data \cite{rusada}. For bias-free results, all trained models were K-fold cross-validated.

\section{Experimental Analysis}

\subsection{Performance Evaluation}
 
The clusters created for analysis were evaluated empirically. Although the analyzed dataset had labels and the results could be evaluated using a metric, this was not done, as evaluating clustering algorithms using labels is not recommended \cite{farber2010using}. 

For the supervised algorithms recall, macro-F1 and Area Under Curve of ROC (AUC-ROC) were used to evaluate the models. Recall is the measure of the fraction of good programmers correctly identified as good programmers \cite{powers2011evaluation}. Recall is calculated as equation (1). F1 is an evaluation metric measured by combining precision and recall, and it is calculated as (3) \cite{powers2011evaluation}. Macro-F1 is the arithmetic mean of the per class F1 scores. It has been selected as an evaluation criterion because the training data was imbalanced, and macro-F1 is a good metric for imbalanced data \cite{wu2016constrained}. AUC-ROC is the area under a ROC curve that allows comparison between models \cite{powers2011evaluation}. Apart from these evaluation metrics, balanced accuracy was also reported. Balanced accuracy is defined as the average of recall obtained on each class \cite{balancedAccuracy}.

\begin{equation} 
Recall = \frac{True\:Positive}{True\:Positive + False\:Negative}
\end{equation}

\begin{equation} 
Precision = \frac{True\:Positive}{True\:Positive + False\:Positive}
\end{equation}

\begin{equation} 
F1 = \frac{2 * Precision * Recall}{Precision + Recall}
\end{equation}
 
\subsection{Results and Discussion}
After analyzing the contest results from 2015, 2016, and 2017, it was discovered that they were similar. Therefore, only the results of one year (2016) are shown in this section. 

\begin{figure*}[t]
\captionsetup{justification=centering}
\begin{multicols}{4}
    \centering
    \includegraphics[width=\linewidth]{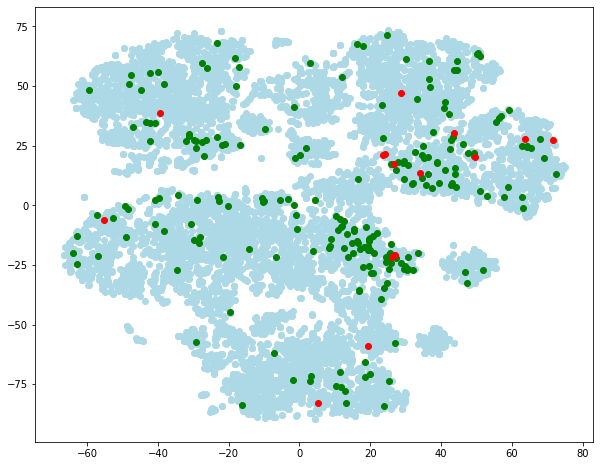}\par
    \subcaption{Qualification round - 565238}
    
    \includegraphics[width=\linewidth]{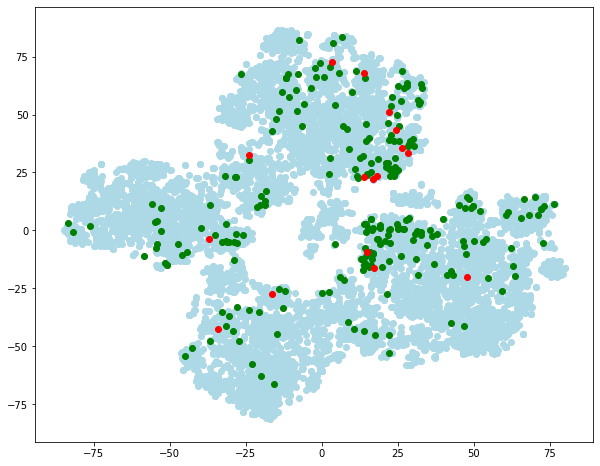}\par
    \subcaption{Qualification round - 563469}
    
    \includegraphics[width=\linewidth]{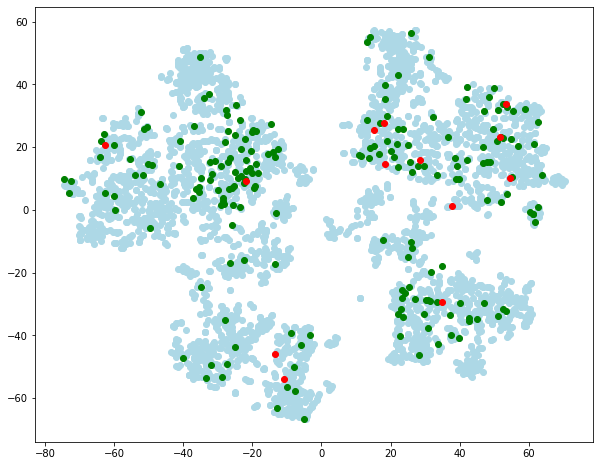}\par
    \subcaption{Qualification round - 563631} 
    
    \includegraphics[width=\linewidth]{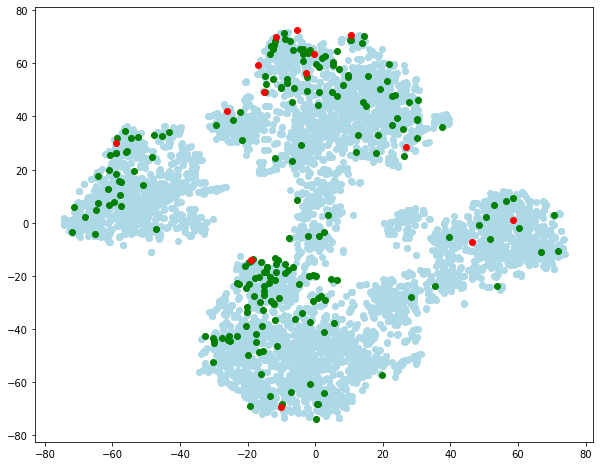}\par
    \subcaption{Qualification round - 573860}
    \end{multicols}
\begin{multicols}{4}
    \includegraphics[width=\linewidth]{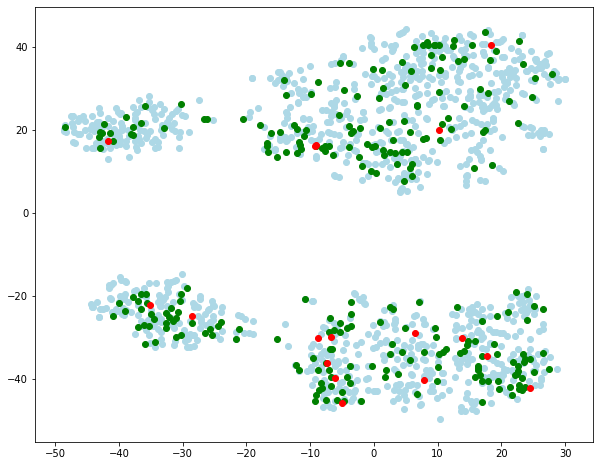}\par
    \subcaption{Round 2 - 567760}
    
    \includegraphics[width=\linewidth]{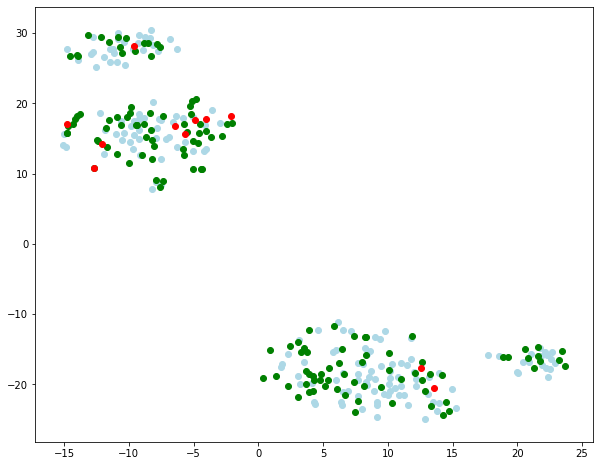}\par
    \subcaption{Round 2 - 572360}
    
    \includegraphics[width=\linewidth]{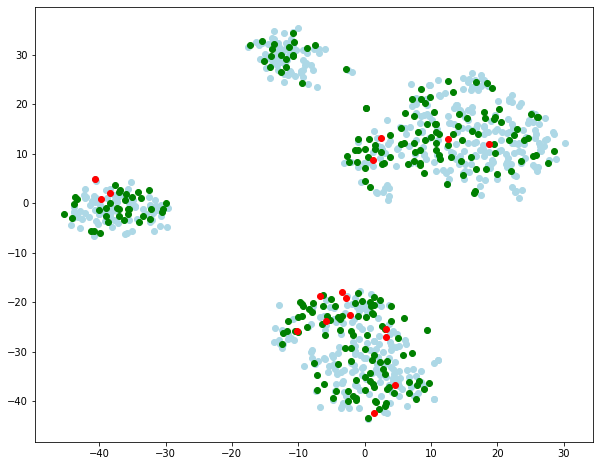}\par 
    \subcaption{Round 2 - 571844}
    
    \includegraphics[width=\linewidth]{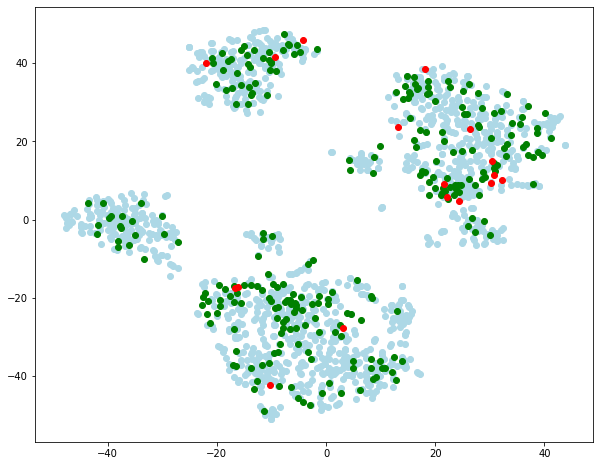}\par 
    \subcaption{Round 2 - 571860}
\end{multicols}
\begin{multicols}{4}
    \includegraphics[width=\linewidth]{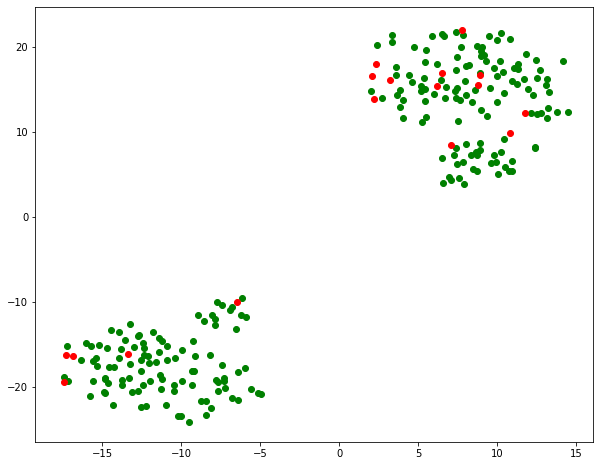}\par
    \subcaption{Round 3 - 574081}
    
    \includegraphics[width=\linewidth]{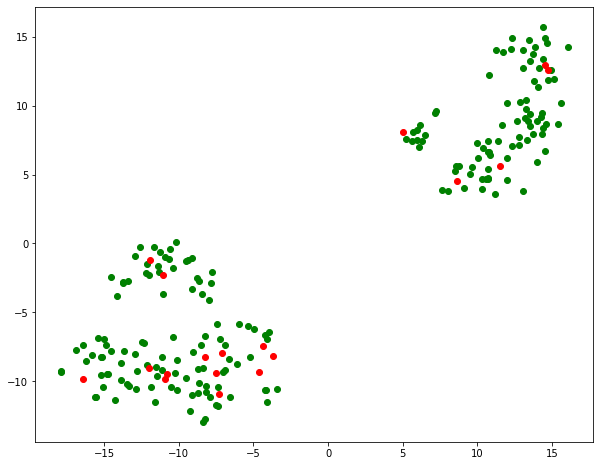}\par
    \subcaption{Round 3 - 563236}
    
    \includegraphics[width=\linewidth]{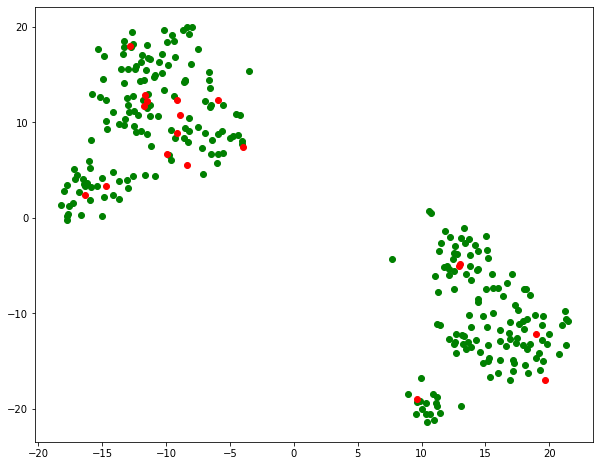}\par 
    \subcaption{Round 3 - 563461}
    
    \includegraphics[width=\linewidth]{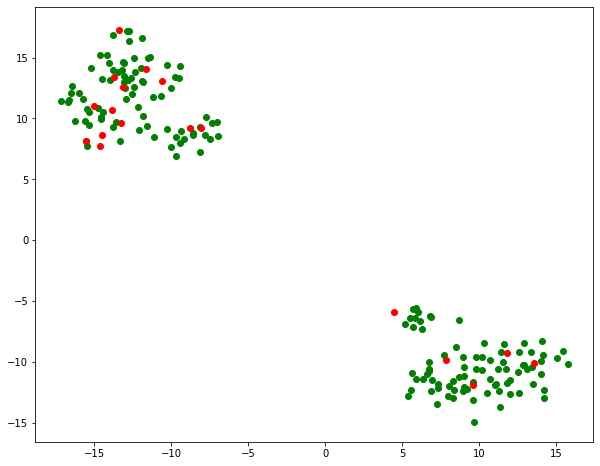}\par 
    \subcaption{Round 3 - 512508}
\end{multicols}
\caption{t-SNE Style Clusters of GCJ 2016}
\label{fig:tsne}
\end{figure*}

Figure \ref{fig:tsne} shows the t-SNE clusters of all the problems in the dataset of the year 2016. The caption of each image shows the round and the problem number. From the graphs, it can be said that 4 stylistic clusters exist for each solution. The most important features of the clusters determined by a Random Forest Classifier are shown in Table \ref{tab:featimportance}. The importance of all other features was less than 0.05. It is seen that \emph{newLineBeforeOpenBrace} and \emph{tabsLeadLines} are the most prominent features in separating the clusters. A manual inspection of the codes also proved the findings to be true. The discovered clusters are formed around the following feature combinations:
\begin{itemize}
    \item new line before opening braces, tabs lead lines
    \item no new line before opening braces, tabs lead lines
    \item new line before opening braces, whitespace lead lines
    \item no new line before opening braces, whitespace lead lines
\end{itemize}

Although style clusters were found, the good programmers were almost equally distributed among them. As a result, we cannot conclude that good programmers belong to a specific cluster. 

\begin{table}[]
\centering
\caption{Feature importance of Clusters}
\label{tab:featimportance}
\begin{tabular}{|l|l|}
\hline
\textbf{Features}      & \textbf{Importance} \\ \hline
newLineBeforeOpenBrace & 0.282               \\ \hline
tabsLeadLines          & 0.163               \\ \hline
numTabs/length         & 0.151               \\ \hline
numSpaces/length       & 0.103               \\ \hline
\end{tabular}
\end{table}



The results of the supervised machine learning models are shown in Table \ref{tab:modelresults}. BRF, LR, SVC and RUSAda performed better than the dummy model, which indicates that some patterns can be identified by the models that can be used to predict good programmers. Also, BRF outperformed all models in terms of Recall, macro-F1 and AUC-ROC. While different studies have used code style for various aspects such as author identification and plagiarism detection, none of the studies have dealt with good programmer identification. Therefore, we cannot compare our results with those of existing studies. However, our results can inspire further research on the relationship between code style and the coding ability of programmers. 

\begin{table}[h]
\centering
\caption{Prediction Results of Supervised Learning Models}
\label{tab:modelresults}
\begin{tabular}{|l|c|c|c|c|}
\hline
\textbf{Model} & \multicolumn{1}{l|}{\textbf{Recall}} & \multicolumn{1}{l|}{\textbf{\begin{tabular}[c]{@{}l@{}}macro-\\F1\end{tabular}}} & \multicolumn{1}{l|}{\textbf{\begin{tabular}[c]{@{}l@{}}AUC-\\ROC\end{tabular}}} & \multicolumn{1}{l|}{\textbf{\begin{tabular}[c]{@{}l@{}}Balanced\\ Accuracy\end{tabular}}} \\ \hline
BRF            & 0.650                                & 0.511                                   & 0.695                                 & 0.645                                                                                     \\ \hline
LR             & 0.641                                & 0.523                                   & 0.692                                 & 0.651                                                                                     \\ \hline
SVC            & 0.601                                & 0.523                                   & 0.689                                 & 0.639                                                                                     \\ \hline
RUSAda         & 0.510                                & 0.50                                    & 0.626                                 & 0.590                                                                                     \\ \hline
Dummy          & 0.485                                & 0.412                                   & 0.499                                 & 0.489                                                                                     \\ \hline
KNN            & 0.469                                & 0.494                                   & 0.593                                 & 0.565                                                                                     \\ \hline
DT             & 0.287                                & 0.525                                   & 0.542                                 & 0.542                                                                                     \\ \hline
RF             & 0.185                                & 0.539                                   & 0.664                                 & 0.537                                                                                     \\ \hline
\end{tabular}
\end{table}

\section{Threats to Validity}

This section presents aspects that may threaten the validity of the study:
\begin{itemize}
    \item \textbf{Internal validity:} The result of our analysis largely depends on the stylistic features that were used. Using other stylistic features may affect the results. However, many existing studies \cite{caliskan2015anonymizing, tereszkowski2022towards} have used these features for their analysis, so they can be relied upon.
    
    \item \textbf{External validity:} The analysis was done on the source files of the GCJ dataset. Therefore, the findings of this study may not be generally applicable to contests in other formats. Furthermore, as only C++ codes were selected for analysis, it cannot be said whether stylistic features of other programming languages will show similar results. Additionally, the criteria for defining a good programmer are subjective and could be defined in other ways depending on the context. In such contexts, our results can not be generalized.
\end{itemize}

To ensure the reliability of the study, the  analysis results are made publicly available in Jupyter notebooks at \href{https://github.com/rafed123/GcjStyleAnalysis}{github.com/rafed/GcjStyleAnalysis}.

\section{Conclusion}
This paper explores whether code style can be used to identify good programmers. The study was conducted on C++ solutions from the Google Code Jam contest. Clustering techniques such as t-SNE and hierarchical agglomerative clustering were used to discover whether style clusters exist and if good programmers could be attributed to any of them. Although four style clusters were found, good programmers could not be associated with a particular cluster. However, supervised machine learning showed that stylistic attributes can be used to predict good programmers. Seven machine learning models were trained and evaluated using recall, macro-F1 and AUC-ROC. A Balanced Random Forest yielded the best results with 0.650 recall, 0.511 macro-F1 and 0.695 AUC-ROC. The results indicate that code style can be used as a measure to identify good programmers.

Future research will examine if defining style guidelines based on the coding style of skilled programmers enhances the quality of software. Additionally, it is possible to investigate how the current recruitment procedures might be efficiently linked with the prediction of good programmers utilizing code style. There is also potential for improving our results using other techniques.

\bibliography{sample-ceur}

\end{document}